\documentclass[twocolumn,showpacs,preprintnumbers,amsmath,amssymb]{revtex4}
\usepackage{graphicx}
\usepackage{dcolumn}
\usepackage{bm}

\begin{document}

\newcommand{\FH}{ferrihydrite\;}
\newcommand{\MAG}{$Fe_3O_4$ \;}
\newcommand{\MOS}{M\"{o}ssbauer\;}

\title{Large magnetic anisotropy in Ferrihydrite nanoparticles synthesized from reverse micelles}

\author{E. L. Duarte, R. Itri, E. Lima Jr.}
\address{Instituto de F\'{i}sica,  Universidade de S\~{a}o Paulo, \\
CP 66318, S\~ao Paulo, 05315-970, Brazil}

\author{M. S. Baptista}
\address{Instituto de Qu\'{i}mica, Universidade de S\~{a}o
Paulo, Av. Prof. Lineu Prestes 748, S\~{a}o Paulo, Brazil.}

\author{T. S. Berqu\'{o}}
\address{Institute for Rock Magnetism,  University of Minnesota,
100 Union Street S.E. Minneapolis, MN 55455-0128 U.S.A.}

\author{and G. F. Goya}
\address{Instituto de Nanociencias de Arag\'{o}n (INA) Universidad de Zaragoza, Pedro Cerbuna 12, (50009),
Zaragoza, Spain.} \email{goya@unizar.es}

\date{\today}

\begin{abstract}
Six-line ferrihydrite(FH) nanoparticles have been synthesized in the
core of reverse micelles, used as nanoreactors to obtain average
particle sizes $<d>$ $\approx$ 2 to 4 nm. The blocking temperatures
$T_B^m$ extracted from magnetization data increased from $\approx
10$ to $20$ K for increasing particle size. Low-temperature \MOS
measurements allowed to observe the onset of differentiated
contributions from particle core and surface as the particle size
increases. The magnetic properties measured in the liquid state of
the original emulsion showed that the \FH phase is not present in
the liquid precursor, but precipitates in the micelle cores after
the free water is freeze-dried. Systematic susceptibility
$\chi_{ac}(\emph{f},T)$ measurements showed the dependence of the
effective magnetic anisotropy energies $E_{a}$ with particle volume,
and yielded an effective anisotropy value of $K_{eff} = 312\pm10$
kJ/m$^3$.

\end{abstract}


\maketitle

\section{\label{sec:INT}INTRODUCTION}

Ferrihydrite is a poorly crystalline Fe$^{3+}$ oxyhydroxide whose
detailed crystal structure has remained elusive up to now, as
reflected by the identification of the known phases as '2-line' and
'6-line' on the basis of their two or six broad X-ray diffraction
(XRD) peaks. It has been proposed that the 2-line type consists of
extremely small crystalline domains, whereas the 6-line has larger
crystal domains of hexagonal structure with unit-cell with
parameters a = 2.96 {\AA}, c = 9.40 {\AA}. \cite{DRI93} More
recently, Jansen \textit{et al}. \cite{JAN02} have refined their XRD
data by considering a space group P 1c phase plus a defective phase
with P3 space group. Electron Microscopy and \MOS spectroscopy have
been employed quite successfully for identifying the properties of
both kind of structures regarding their local structural and
magnetic properties. \cite{MUR96,MUR80,JOL04,PUN04} These works on
\FH have consistently provided a landscape where some aggregates
with ordered structure, with small regions of $\approx 2-3$ nm,
display definite lattice fringes and narrow distribution of
hyperfine parameters. This iron oxyhydroxide phase is found in many
biological systems (e.g., iron-reducing bacteria, \textit{Aedes
aegypti} mosquito), as well as in hydrometallurgical operations (as
an undesired precipitate). \cite{UDOFE} Along with the interest of
the basic physical mechanisms governing structure and surface
formation, \FH has also significance because it constitutes the core
of ferritin, a protein that plants and animals use to sequester and
store iron, providing a fully biocompatible material to carry iron
particles in potential drug delivery applications. \cite{PAR03}

\begin{figure}
\caption{\label{fig:micelles} Schematic diagram showing the
reverse-micelles synthesis of \FH: the two reverse micelles
solutions (a) with A and B contents of their aqueous phases are
mixed (b) and stirred, to exchange aqueous material and form AB
solution. The surfactant-to-water ratio W controls the micelle
diameter, and therefore the final particle sizes (c).}
\includegraphics[width=7.9cm]{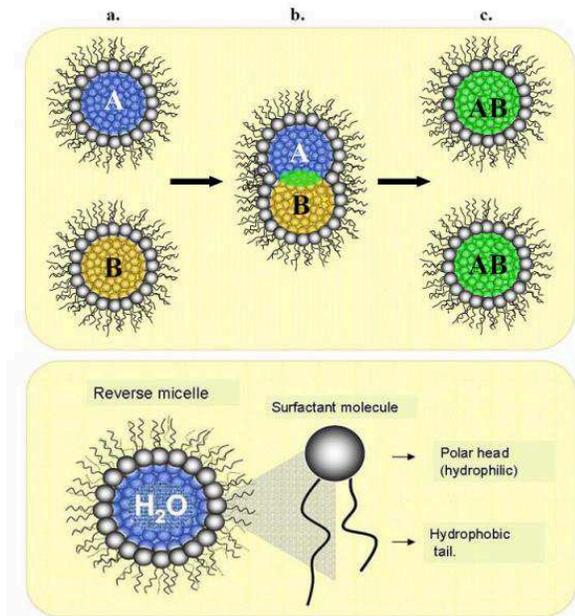}
\end{figure}

Regarding the magnetic properties of \FH, many reported studies on
its behavior are based on its biological complex ferritin, using
either $^{57}$Fe \MOS spectroscopy \cite{BEL84,BAU89} or
magnetization measurements \cite{FH8,FH9}. The main results point to
a cluster-like structure with $Fe^{3+}$ ions antiferromagnetically
ordered within the core and a small uncompensated magnetic moment
(probably at the particle surface). But the actual physical units
from which the observed cluster or superparamagnetic (SPM) behavior
is originated have not been univocally defined.

\begin{figure}
\caption{\label{fig:XRD} X-ray powder diffraction data from the four
samples synthesized using reverse micelles at $W=10, 15, 20$ and
$30$. The peak positions of two-line (full arrows) plus the
additional for six-line (open arrows) ferrihydrite structure are
shown at the top of the figure.}
\includegraphics[width=7.9cm]{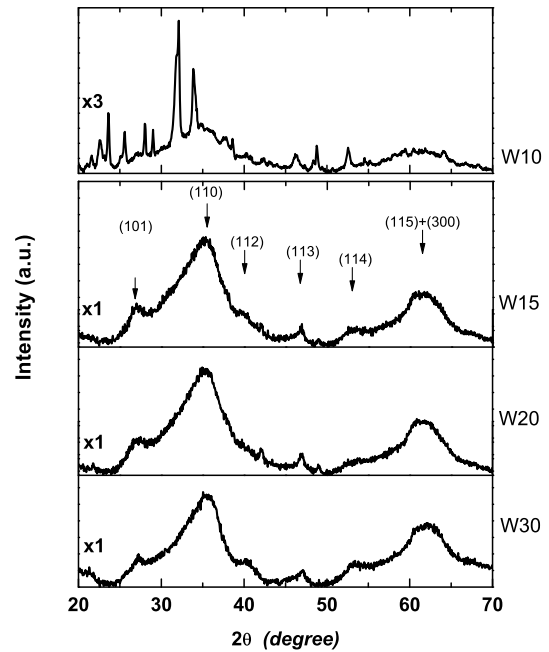}
\end{figure}

For antiferromagnetic (AFM) particles of few-nanometer diameter, a
small uncompensated magnetic moment is expected to arise from
defects within the particle and/or from the unpaired surface moments
of the particle. \cite{NEE61} The net magnetic moment of the
particle reverses among different spatial orientations with a
characteristic time $\tau$ that depends on temperature and particle
volume. Within N\'{e}el's model \cite{NEE49} the probability of
switching the particle's magnetic moment is a thermally activated
process described by

\begin{eqnarray}
\tau=\tau_{0}\exp\left(-\frac{E_a}{k_{B}T}\right),\label{eq:arrhenius1}
\end{eqnarray}

where $E_a$ is the energy barrier that separates two energy minima
between magnetization states (up and down), and $k_B$  is the
Boltzmann constant. The pre-exponential parameter $\tau_{0}$ is
usually assumed to be of the order of $10^{-9}$ $-$ $10^{-11}$ $s$,
and dependent on the reversal mechanism.\cite{MOR90A,DOR97}

\begin{figure}
\caption{\label{fig:TEM} TEM micrograph recorded from sample
$W=15$.}
\includegraphics[width=4.0cm]{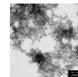}
\end{figure}

In this work we present a characterization of \FH particles with
different average diameters ranging from \emph{ca.} $2$ to $4$ nm,
synthesized using reverse micelles as nanoreactors with the aim to
have a particle volume control. Reverse Micelles (RM) consist of
nanometer-sized aqueous droplets suspended in a nonpolar continuous
organic phase by surfactant shells. The size of the water cores is
determined by the water-to- surfactant molar ratio, W. Some aqueous
phase reactions, including chemical precipitation of solid
particles, can be accomplished in the micelle core. In particular,
RM has been successfully employed as a nanoreactor in the synthesis
of some ferrites (magnetite, $CoFe_{2}O_{4}$, $ZnFe_{2}O_{4}$,
$MnFe_{2}O_{4}$ among others) by several
groups.\cite{PIL00,CON99,LIU00} However, after the precursor is
formed in the micellar core, it is extracted and submitted to a
thermal treatment to reach the desired compound with diameters
ranging from $10$ to $20$ nm. In our case, we have realized that
such precursor is, indeed, made up of 6-L ferrihydrite nanoparticles
with size diameters smaller than $4$ nm depending on W. To the best
of our knowledge, it is the first time that the magnetic properties
of this kind of ferrihydrite synthesized in the core of RM is
reported. The main objective is to investigate the contribution of
the finite size and surface effects on the magnetic properties
through their evolution for increasing particle size.

\section{\label{sec:EXP}MATERIALS AND METHODS}

\subsection{\label{sec:EXP1}Chemicals and Synthesis}

Surfactant sodium \textit{bis(2-ethylhexyl)} sulfosuccinate (AOT)
was purchased from Fluka Chemicals; isooctane was from Synth, tetra
hydrated ferrous chloride ($FeCl_2.4H_2O$) and sodium hydroxide
(NaOH) were from Merck. All reagents were used as received. Water
was distilled in all-glass apparatus, filtered and de-ionized
(milli-Q water). Ethanol and acetone were used during the extraction
process of nanoparticles from reverse micelles. All experiments were
performed at room temperature of $23(1)^{\circ}C$.

The protocol for synthesizing nanoparticles in a micellar
environment is well known in the literature \cite{CAR99,MOT95}, and
consists in mixing two reverse micelles ($RM$) solutions with the
same surfactant composition, but differing in the content of their
aqueous phases, as shown schematically in figure \ref{fig:micelles}.
In our case, both micellar solutions were composed of $0.4M$ AOT in
isooctane, one of them containing 0.05 M aqueous solution of
$FeCl_2$ and the other 0.20 M of sodium hydroxide aqueous solution.
For the mixture, 200 $ml$ of the $RM$ solution containing $FeCl_2$
was vigorously stirred, whereas 200 $ml$ of the $RM$ solution
containing $NaOH$ was drop wised. The final solution was further
stirred for two hours to complete the homogenization. The solution's
color changes from bright yellow to dark brown indicating the
formation of iron-oxide particles \cite{CON99,LIU00}. Four different
samples were synthesized, labeled by the molar ratio between AOT and
the aqueous concentration as $W =$[water]/[AOT], with W values of
$10, 15, 20$ and $30$ (see Table \ref{tab:param}). In order to
extract the synthesized iron oxide nanoparticles from the $RM$, the
$isooctane$ was firstly evaporated by means of a rotoevaporator.
Then, the sample was several times washed with ethanol to remove the
surfactant by centrifuging. Finally, the samples were washed with
acetone to improve the drying. All samples were stored at room
temperature and low pressure in a camera containing silica-gel.
Elemental analysis indicated that the resulting dried powders
contained c.a.$12 wt \%$ of C in their composition. The presence of
C atoms in the synthesized sample is due to the surfactant residues
not well eliminated in the washing process, that might remain
attached to the nanoparticle surface.

\subsection{\label{sec:EXP3}Apparatus}

X-ray diffraction ($XRD$) was used to investigate the
crystallographic structure of the powder samples. The experiments
were performed by means of a Rigaku-Denki powder diffractometer with
a conventional X-ray generator ($Cu$ $K_\alpha$ radiation, $\lambda
=1.5418$ {\AA} and a graphite monochromator) coupled to a
scintillation detector. The angular scanning performed on all
samples ranged from $20^{o}$ up to $70^{o}$ with $0.05^{o}$
step-widths. The average particle diameter ($d_{XRD}$) was
calculated from the full width at half maximum of the more intense
reflection using Scherrer's equation \cite{SCH}.

\begin{figure}[t]
\caption{\label{fig:mosrt} \MOS spectra taken at room temperature
for samples W10 to W30. Open symbols are the experimental points,
and solid lines the fitted spectra.}
\includegraphics[width=7.9cm]{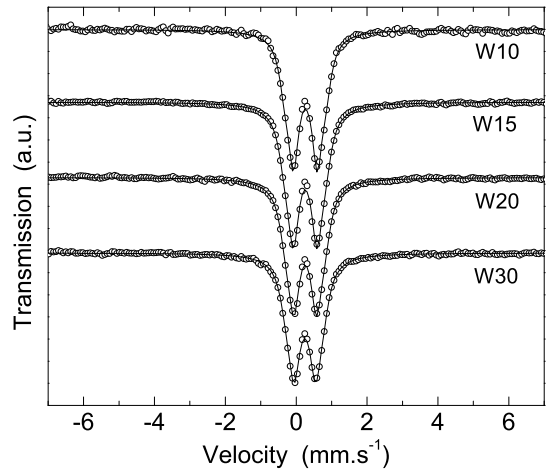}
\end{figure}

M\"ossbauer spectroscopy (MS) measurements were performed between
$4.2$ and $296$ $K$ in a liquid He flow cryostat, with a
conventional constant-acceleration spectrometer in transmission
geometry. The spectra were fitted to Lorentzian line shapes using a
non-linear least-squares program, calibrating the velocity scale
with a foil of $\alpha-Fe$ at $296$ $K$. When necessary, a
distribution of hyperfine magnetic fields, isomer shift and
quadrupole splitting have been used to fit the spectra.
Magnetization and ac magnetic susceptibility measurements were
performed in a commercial SQUID magnetometer both in
zero-field-cooling (ZFC) and field-cooling (FC) modes (always in the
heating direction) between $1.8$ $K$ $<$ $T$ $<$ $250$ $K$ and under
applied fields up to $9$ $T$. The frequency dependence of both
in-phase $\chi'(T)$ and out-of-phase $\chi''(T)$ components of the
ac magnetic susceptibility were measured by using an excitation
field of $1 - 4$ $G$ and driving frequencies $0.01$ $Hz$ $\leq$ $f$
$\leq$ $1500$ $Hz$.

\section{\label{sec:RESULTS}EXPERIMENTAL RESULTS}

\subsection{\label{sec:A}Structural Analysis and Particle Dimension}

\begin{figure}
\caption{\label{fig:mosdist} Fitted \MOS spectra at $T = 4.2$ K for
samples W10 to W30 using a distribution of fields.  Open symbols are
the experimental data, and solid lines the fitted spectra. The bars
indicate the amount of resonant effect (percent) for each spectrum.
The right column shows the resulting histogram for the obtained
hyperfine field $B_{hyp}$.}
\includegraphics[width=7.9cm]{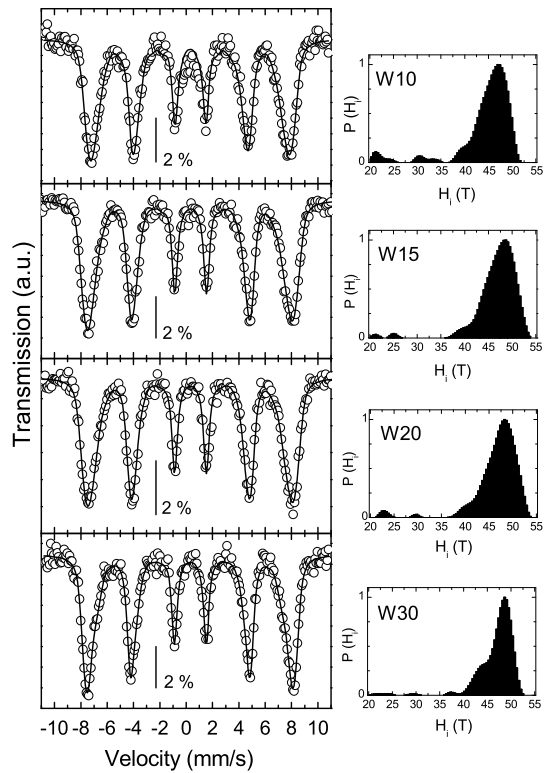}
\end{figure}

X-ray profiles performed on dried powders showed two prominent peaks
at $2\theta\sim35.5^{o}$ and $\sim61.5^{o}$, as indicated by the
arrows in the figure \ref{fig:XRD}, corresponding to q values of
2.49 {\AA}$^{-1}$ and 4.17 {\AA}$^{-1}$, respectively. The broad
peak at $\sim61.5^{o}$ seems to be two peaks combined. Additionally,
small peaks at $\sim41.1^{o}$ and $\sim53.7^{o}$ can be noticed, and
also a very small feature at $\sim47^{o}$. These features on the XRD
pattern agrees with a 6-line ferrihydrite phase. \cite{JAM98} The
sample $W10$ presents additional peaks that are identified as
diffraction peaks from $Na_{2}SO_{4}$ crystals. This phase is likely
to be formed during the AOT extraction process, since it is known
that at $W10$ values all water in the pool is strongly bounded and
there is a high local amount of hydroxyl ions (related to the AOT
hydrolysis process). When a nanoparticle grows in this environment,
it must also be strongly bound to the hydrolyzed AOT polar head,
yielding $Na_{2}SO_{4}$ formation on the nanoparticle surface. This
is not observed at concentration ratios $W15$ to $W30$ because the
presence of bulk water facilitates the nanoparticle solvation and a
decrease in concentration of local hydroxyl ions.

Because of the low crystallinity of ferrihydrites, the broad peaks
cannot be directly used for an accurate determination of the
particle size through the Scherrer's formula, so that further
transmission electron microscopy (TEM) images were used to better
define the particle dimension. Figure \ref{fig:TEM} shows as an
example a TEM micrograph from W15 sample. All images were recorded
with a slow scan CCD camera (Proscan) and processed in the AnalySis
3.0 software that allows to work with enhanced images in such a way
that particle-by-particle size measurement can be performed with
high precision. Table \ref{tab:param} contains the average particle
diameter $d_{XRD}$ obtained from the diffraction data, calculated
from the broad diffraction peak at $2\theta\sim35.5^{o}$. The
resulting values were compatible with the mean values observed in
electron microscope images and expected for six-line ferrihydrites.
\cite{UDOFE,JAM98} Note that the average size seems to increase from
$W10$ to $W30$, suggesting that the micelle pool features might, in
our case, define the iron-oxide based nanoparticle size.

\begin{table*}

\caption{\label{tab:param}Magnetic and hyperfine parameters for
samples with average grain sizes $\left<d\right>_{XRD}$ as estimated
from XRD data: blocking temperature $T_B^{m}$; coercive field $H_C$,
energy barrier $E_a^{zfc}$, and energy barrier $E_a^{ac}$. Hyperfine
parameters from the distribution fit of figure \ref{fig:mosdist}:
Isomer Shift (IS); hyperfine field at the distribution maximum
($B_{max}$); average hyperfine field ($B_{mean}$) and variance
$\mathfrak{V}$ of the fitted distribution.} \footnotesize\rm
\begin{tabular*}{\textwidth}{@{}l*{15}{@{\extracolsep{0pt plus12pt}}l}}
Sample&$\frac{[H_{2}O]}{[AOT]}$&$<D>_{XRD}$& $T_{B}^{m}$ & $E_a^{ZFC}$ &  $H_C(5K)$ &  $E_a^{ac}$  &  IS   &  $B_{max}$  &  $B_{mean}$  &   $\mathfrak{V}$  \\
&  & (nm) & (K)$^{a}$ & $(\times 10^{-21}$ J)$^{a}$ & (kOe) &
$(\times 10^{-21}$ J)$^{b}$ &
(mm/s) & (T) & (T)  &  (a.u.) \\

W10&10&2.4(1)&10.4(5)&4.5(3)&1.23(9)&3.1(3)&0.42(1)&47.1(1)&44.5&3470\\
W15&15&3.2(1)&17.4(5)&7.8(4)&1.22(9)&5.0(4)&0.42(1)&48.3(1)&46.6&2336\\
W20&20&3.1(1)&20.1(5)&10.1(3)&1.22(9)&6.6(3)&0.43(1)&48.5(1)&46.8&2004\\
W30&30&3.9(1)&29.5(5)&16.5(5)&1.13(9)&11.7(7)&0.41(1)&48.7(1)&46.9&1706\\

\end{tabular*}
$^{a}$ From ZFC curves, see text and fig \ref{fig:figzfcfc}.\\
$^{b}$ From $\chi(f,T)$ data.
\end{table*}

\subsection{\label{sec:C}\MOS Measurements}

In the SPM state the magnetic hyperfine interactions are averaged to
zero due to fast relaxation of the particle magnetic moments, so
that the resulting \MOS spectra consist of a paramagnetic-like
doublet. Accordingly, the room temperature spectra recorded for
samples W10 to W30 (Figure \ref{fig:mosrt}) showed the expected
doublet, with essentially identical hyperfine parameters, i.e.,
quadrupole splitting QS $= 0.60-0.63(2)$ mm/s and Isomer Shift IS $=
0.24-0.25(1)$ mm/s for all samples. In the blocked state, the
nuclear spin levels of the $^{57}Fe$ probe are split by the magnetic
interactions yielding a six-line \MOS spectrum. For a few-nanometer
particle, it is known that surface contributions become increasingly
important, resulting in a experimentally observed decrease of
hyperfine magnetic fields $B_{hyp}$, assigned to weaker exchange
fields sensed by the atoms at the surface. \cite{RUS03} For the
present particles with average diameters $<d> \approx 2.4 - 4.0$ nm
(corresponding to few lattice parameters, respectively), about
$75-80 \%$  of the spins will be located at the surface of the
particle. Accordingly, the resulting spectra at T = 4.2 K (fig.
\ref{fig:mosdist}) were composed of an unresolved distribution of
magnetic sextets reflecting the different local Fe environments and,
therefore, hyperfine field distributions were used for fitting the
low temperature spectra. The resulting parameters (Table
\ref{tab:param}) showed a reduced value of hyperfine fields (both
the maximum $B_{max}$ and mean $B_{mean}$ values) for the W10 sample
that increase smoothly for the larger particle sizes, reflecting the
large surface effects of the smallest particles. For samples W15,
20, 30 the hyperfine parameters are in agreement with 6-lines
ferrihydrite though with a slightly reduced $B_{max}$ when compared
to the corresponding values in "well-crystallized" 6-line
ferrihydrite. \cite{THE06} This $B_{max}$ decrease could be related
to lattice defects (vacancies). It is possible to observe that the
second moment $\mathfrak{V}$ of the distribution, which describes
how the fields are distributed around $B_{mean}$, steadily decreases
with increasing particle size. For sample W30 the distribution is
quite narrow, and the profile shows an incipient shoulder at $B
\approx 44$ T. This shoulder suggests a difference between surface
and core spin environments for the largest particles. For sample
$W20$ this shoulder is still noticeable, but for smaller particles
($W15$ and $W10$) the spin environments of surface and core become
increasingly alike, the corresponding hyperfine fields merge each
other and the resulting distributions become broader.

\begin{figure}
\caption{\label{fig:figzfcfc} Temperature variation of magnetization
for H=$100$ Oe in field-cooled (FC) and zero-field-cooled (ZFC)
conditions.}
\includegraphics[width=7.9cm]{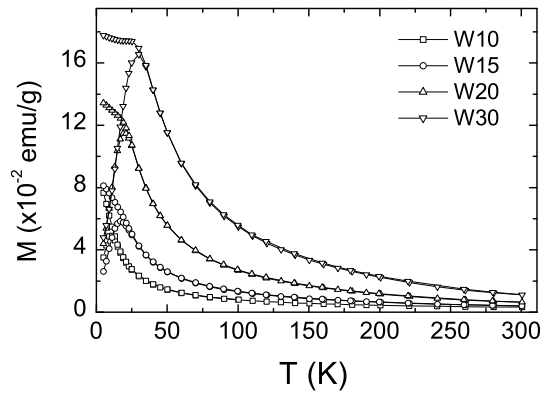}
\end{figure}

\subsection{\label{sec:B} Magnetization data}

\begin{figure}
\caption{\label{fig:wliqzfc} Magnetization curves in ZFC modes of
the liquid precursor \textit{as prepared} (W$_{liq}$) and
lyophilized (W$_{gel}$). The inset shows a magnification of the ZFC
peak for W$_{gel}$, indicating long-range magnetic order.}
\includegraphics[width=7.9cm]{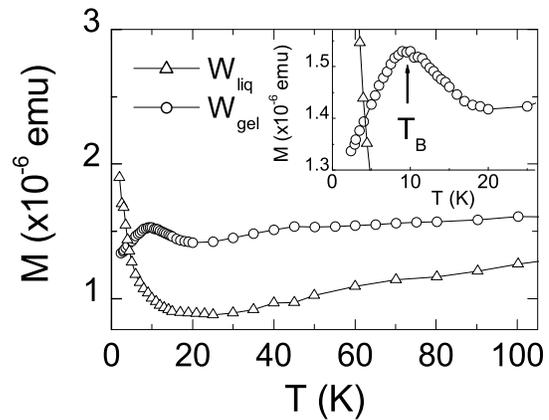}
\end{figure}

The magnetization curves M(T) taken in field-cooling (FC) and
zero-field-cooling (ZFC) modes with $H = 100$ Oe are shown in Fig.
\ref{fig:figzfcfc}. The maxima of the ZFC branches, from which the
blocking temperature $T_B^{m}$ is usually defined, shifts smoothly
to larger temperatures for increasing \texttt{W} values (Table
\ref{tab:param}). It is well known that interparticle interactions
can influence the blocking process and even the magnetic dynamics of
the system \cite{DOR97}. The interparticle interactions are mainly
determined by the ratio of the average distance between particles to
their size. For the present concentrated (powder) samples dipolar
interactions (DI) are likely to exist and to influence to some
extent the magnetic behavior, although we are unable to quantify
these parameters. However, the fact that W10 and Wgel samples, with
identical particle size but different interparticle distances shown
nearly identical blocking temperatures (fig \ref{fig:figzfcfc} and
\ref{fig:wliqzfc} and discussion below) suggest that DI are not the
dominating contribution to the energy barrier in these particles.
For each sample, both ZFC and FC curves (measured for increasing
temperatures) merge each other at temperatures $T_{rev}$ very close
($\approx 2$ K) above $T_B^m$, and the system displays full
reversibility above this temperature. In real systems the
distribution of particle sizes yields a distribution of $T_B$, so
that $T_{rev}$ can only be reached when all particles, including the
largest ones, become unblocked. The close values between $T_B^m$ and
$T_{rev}$ is a clear indication of the extremely narrow distribution
of particle sizes in our samples.

In order to see whether the magnetic phase is formed inside the
reverse micelles in the liquid state, we have measured the $M(T)$
curves for the precursor aqueous solutions. Figure \ref{fig:wliqzfc}
shows the results for the precursor liquid having molar ratio $W =
10$ (i.e., identical to the W10 sample). The observed paramagnetic
behavior down to $T=1.8$ K demonstrates that no formation of phases
with long-range magnetic order occurs at the liquid stage of the
synthesis process. To further investigate the formation process, the
original solution was liophylized at $P \approx 10^{-4}$ Torr during
1 h, until most of the liquid was freeze-dried. The resulting ZFC
curve of $W_{gel}$ (fig. \ref{fig:wliqzfc}) shows a maximum in the
ZFC branch, clearly showing the development of the magnetic phase
having large-range magnetic order. The position of this maximum,
located at the same temperature $T_B^m = 10.0(5)$ K than the
corresponding W10 sample, indicates that particles with the same
average particle size are present in both $W_{gel}$ and W10 samples.
The above results support the idea that the particle growth is
limited by the micelles, whose boundary determines the final
particle volume.

\begin{figure}
\caption{\label{fig:figMHLT} Magnetization of the W10 sample at $T=
4.2$ K and $10$ K showing the large coercive field and a non
saturation up to 9 T.}
\includegraphics[width=7.9cm]{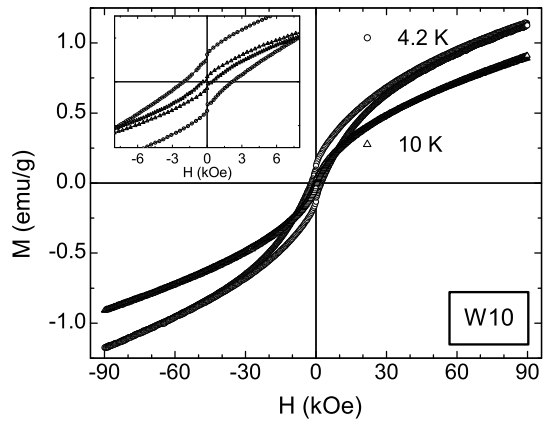}
\end{figure}

The magnetization as a function of applied fields M(H) taken at $T
=$ 4.2 K showed non saturating behavior up to $H = 9 T$ (see fig.
\ref{fig:figMHLT}), with coercive field values $H_C \simeq 1.2$ kOe,
which are compatible with previously reported values for this
material. \cite{THE06} This non-saturating behavior has been
explained by N\'{e}el as originated in the contribution of the core
susceptibility $\chi_a$ of the antiferromagnetic (AFM) fine
particles.\cite{NEE61} The coercivity of the particles was found to
vanish very close from T$_B$ of each sample, as observed in fig.
\ref{fig:figMHLT} for sample W10, showing almost no coercivity
already at T $= 10$ K.

\subsection{\label{sec:D}Susceptibility measurements}

\begin{figure}
\caption{\label{fig:ac1} Typical set of the real component of ac
susceptibility data $\chi'(f,T)$ as a function of temperature at
different applied frequencies (upper side) and the corresponding
imaginary component $\chi''(f,T)$ (bottom side).}
\includegraphics[width=7.9cm]{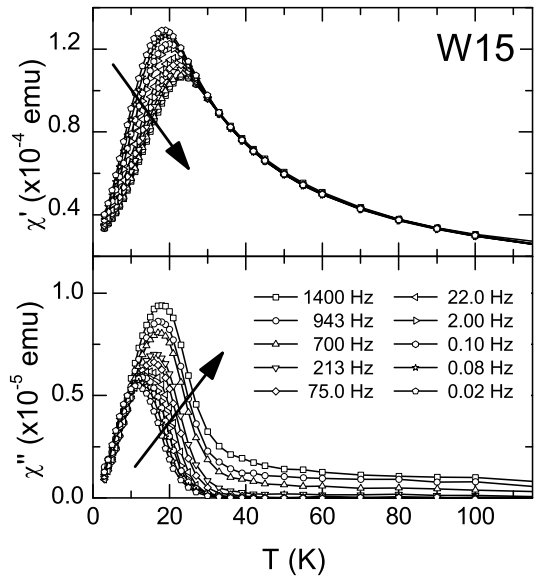}
\end{figure}

A typical set of data from the real  $\chi'$($f$,T) and imaginary
$\chi''$($f$,T) components of the ac susceptibility is shown in
Figure \ref{fig:ac1}. Similar trends were observed for all samples.
Both components of $\chi$($f$,T) exhibit the expected behavior for
SPM systems, i.e., the occurrence of a maximum at a temperature
$T_m$ that shifts towards higher values with increasing frequency.
\cite{DOR97} It can be observed from Table \ref{tab:param} that the
$E_a^{ac}$ values increase from $W10$ to $W30$ samples. The observed
increase of $E_a$ could reflect either the influence of the particle
volume and/or anisotropy on the effective activation energy $E_a$ =
$K_{eff}V$, since our results clearly show (see fig.
\ref{fig:arrheniusall}) that the N\'{e}el-Arrhenius model correctly
suits the behavior of all samples. However, although the relaxation
times $\tau$ of all samples exhibit exponential dependence on
temperature (Eq.\ref{eq:arrhenius1}), the fitted values of $\tau_0$
are increasingly smaller than the $\tau_0 \approx 10^{-9}$ to
$10^{-11}$ s expected for SPM systems. Both effects (i.e., increase
in energy barriers and small relaxation times) are known to occur in
system of interacting particles \cite{DOR97,YOENDOJAP} where, in
addition to the contributions from the intrinsic particle anisotropy
to $E_a$ such as shape, magnetocrystalline, or stress ansotropies,
interparticle interactions (dipolar or exchange) can also modify the
energy barrier.

The fitting of the experimental $T(f,H=0)$ data using eq.
\ref{eq:arrhenius1} and the average particle radii from XRD data
yielded the values of $K_{eff} = 312\pm10$ kJ/m$^3$, which lays
within the wide range of values reported for the effective
anisotropy constant of synthetic \FH $K_{eff}^{bulk}$ = $35-610$
$kJ/{m^3}$. \cite{SUZ96,GIL00,HAR99}

\section{\label{sec:DISCU}DISCUSSION}

Single-domain magnetic particles reverse their magnetization
direction due to thermal agitation, with a characteristic time
$\tau$ and, therefore, the temperature of the blocking transition
depends on the window time $\tau_M$ of the experimental technique
used. Typical magnetization measurement assumes $\tau_M^{m}$ as
$\approx10^1-10^2$ s, and \MOS spectroscopy $\tau_M^{M}\approx
10^{-8}$ s. A first estimation of the activation energy $E_a =
K_{eff} V$ was made using the values of $T_B^{m}$ and $\tau_M^{m}=
10^2$ and the relation at the SPM transition \cite{DOR97}

\begin{eqnarray}
ln\left(\frac {\tau}{\tau_0}\right)\approx 30 = \frac{E_a}{k_B T_B}
\label{eq:taudc}
\end{eqnarray}.

\begin{figure} [t]
\caption{\label{fig:arrheniusall} Blocking temperature $T_{B}(\ln
f,H)$ dependence from $\chi''(f,T)$ data at $H$ $=$ $40$ $Oe$. The
values of $T_B$ extracted from \MOS measurements are also
plotted.The solid lines represent the best fit to eq.
\ref{eq:arrhenius1}.}
\includegraphics[width=7.9cm]{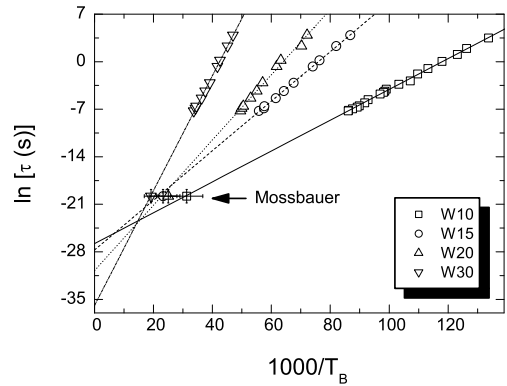}
\end{figure}

The resulting $E_a^{ZFC}$ values are included in Table
\ref{tab:param}. The observed increase of $E_a^{ac}$ could be
attributed to the increasing particle volume on the energy barriers
of the system. It is known that the ratio between the average
distances and sizes of the particles usually determines how much the
interaction influences $T_B$. For example, at low concentrations the
interaction is weak and therefore the single-particle anisotropy is
dominant. However, as already discussed for ac susceptibility data
these values are obtained in non-diluted samples and thus they
include the (unknown) contribution from magnetic dipolar
interactions to the effective anisotropy constant $K_{eff}$. Since
the samples are in the concentrated regime, small changes in
particle volume could affect the dipolar interactions because
particle diameters are comparable to the average distance between
particles. The observed variation of $\tau_0$ in fig.
\ref{fig:arrheniusall} seems too large to be explained only by
volume changes, so that changes in the interparticle interactions
cannot be discarded. The value obtained from eq. \ref{eq:taudc}
amounted to $K_{eff} \approx 320-380$ kJ/m$^3$, which is in
agreement with the $K_{eff} = 312\pm10$ kJ/m$^3$ value from ac
susceptibility discussed above. These values are nearly two orders
of magnitude larger than the usual values for iron oxides. We
propose that this large value for the magnetic anisotropy is mainly
related to the \FH phase, and give some arguments below.

The fit of the frequency dependence (fig. \ref{fig:arrheniusall})
gave very close values of magnetic anisotropy for all samples,
suggesting that the slight spread in $K_{eff}$ values are related to
experimental sources of error. Although this value obtained for the
present non-diluted system also includes the effects of dipolar
interactions, such a large value suggests other sources of
anisotropy. Shape anisotropy is not expected to be the main
contribution due to the spherical morphology observed from TEM
images. We have estimated the shape contribution expected for
deviations into a (e.g. prolate) spheroid with $r$ $=$ $c/a$ ($a$
$=$ $b$ and $c$ are the minor and major axis, respectively), to the
effective anisotropy. For the above situation, the shape-anisotropy
constant given by \cite{KSHAPE}

\begin{eqnarray}
K_{shape}=\pi M^{2}[1-\frac{3}{B^{2}}(\frac{r\ln(r+B)}{B}-1)]
\label{eq:kshape}
\end{eqnarray}

where $M$ is the particle magnetization and $B=\sqrt{r^{2}-1}$. An
upper limit for this contribution can be estimated setting $r$
$\approx$ $2$, and noting that particles with this ratio value, that
should be easily observable, were not detected from TEM images. Even
if particles having this shape were present, they would contribute
to $K_{shape}$ with $\leq$ $75$ $kJ/m^{3}$, which is only $25 - 30
\%$ of the experimental value. It should be noted that the \FH
particles retain their surfactant layer after lyophilization, and
therefore exchange coupling between particles is not expected. We
have discarded possible anisotropy contributions from the organic
coating, since the value of the effective anisotropy is expected to
decrease for increasing particle size when surface effects are
dominant.

It can be noted from figure \ref{fig:arrheniusall} that the linear
fits using eq.\ref{eq:arrhenius1} extrapolate to increasingly large
values at $T_B^{-1}=0$ (i.e., $\tau_0$) for increasing $<d>$ values,
in agreement with the expected dependence of this parameter on
temperature and particle volume. \cite{NOW05} However, a
quantitative analysis of this dependence would require a general
expression for $\tau_0(V,T)$, which is not yet known. \cite{DOR97}.

In conclusion, we have succeeded in controlling the resulting
particle sizes of \FH particles by varying the water and surfactant
molar ratio in the reverse micelle used as a nanoreactor. The
resulting magnetic behavior reflects this control of average
particle size through the smooth variation of the magnetic
parameters, such as blocking temperature $T_B$ and activation energy
$E_a$. By investigating the thermally activated nature of the
blocking process through dynamical data, a rather large anisotropy
constant of $K_{eff} = 320-380$ kJ/m$^3$ has been obtained. The
origin of this anisotropy, which is $\sim 50$ times larger than the
typical for iron oxides, remains uncertain, as well as question
whether it is located at the core or surface of the particles. The
well-known ability of \FH to transform quite easily into other iron
phases, depending on the environmental conditions \cite{JAM98,EGG88}
(further favored in nanoparticles due to the large surface area) led
us to search for evidence of phase evolution in our particles. After
repeating both magnetization and \MOS experiments several weeks
later, we obtained essentially the same results as in fresh samples.

\section{Acknowledgments} This work was supported in part by the
Brazilian agencies FAPESP, CNPq and CAPES/Procad. ELD was a
recipient of CNPQ PhD fellowship. TSB was supported by National
Science Foundation (NSF) grant EAR 0311869 from the Biogeosciences
program. This is IRM publication No. 0604. ELJ is thankful to the
VolkswagenStiftung, Germany, for providing financial support through
a Postdoctoral Fellowship. GFG acknowledges financial support from
the Spanish Ramon y Cajal
program.\\

\bibliography{unsrt}

\end{document}